\documentclass[slac_one]{revtex4}
\usepackage{graphicx}
\usepackage{fancyhdr}
\newcommand{\Niezurawski}{Nie\.zurawski}
\newcommand{\Zarnecki}{\.Zarnecki}

\def\figheight{0.55\textwidth} 
\def\twofigheight{0.5\textwidth} 

\def\eg{\textit{e.g.\ }}
\def\ie{\textit{i.e.\ }}
\def\etal{\textit{et al.}}

\def\bc{\begin{center}}
\def\ec{\end{center}}

\def\bmp{\begin{minipage}}
\def\emp{\end{minipage}}

\def\ar{\rightarrow}
\def\ga{\gamma}
\def\gaga{\ga\ga}

\def\qbar{\bar{q}}
\def\qqbar{q\qbar}

\def\bbar{\bar{b}}

\def\bbbar{ b\bbar }

\def\gagaqq{ \gaga \ar \qqbar }

\def\WW{W^{+} W^{-}}

\def\gagaWW{\gaga \ar \WW}

\def\Wgaga{W_{\gaga}}

\def\epem{ e^{+} e^{-} }

\def\Mh{M_h}

\newcommand{\gagahad}{\( \gaga \ar  \mathit{hadrons} \)}

\newcommand{\btagging}{\( b \)-tagging}

\def\AO{A}
\def\HO{H}

\def\MAO{ M_{\AO} }
\def\MHO{ M_{\HO} }

\def\MAOeq{$ \MAO = $ }

\def\AOHO{ \AO,\HO }
\def\MAOHO{ M_{\AOHO} }

\def\AHbb{ \AOHO \ar \bbbar }
\def\Abb{ \AO \ar \bbbar }
\def\Hbb{ \HO \ar \bbbar }

\def\gagaAbb{ \gaga \ar \Abb }
\def\gagaHbb{ \gaga \ar \Hbb }
\def\gagaAHbb{ \gaga \ar \AHbb }

\def\sgagaAHbb{ \sigma (\gagaAHbb)}

\def\tanb{\tan \beta}
\def\tanbeq{$\tanb = $ }
\def\tbseven{\tanb = 7}

\def\HOgaga{\HO \ar \gaga}

\def\HObb{\HO \ar \bbbar}

\def\BrAOHObb{{\rm BR}(\AO / \HObb)}
\def\BrAOHOgaga{{\rm BR}(\AO / \HOgaga)}

\def\tautau{\tau^{+}\tau^{-}}
\def\gagatautau{\gaga \ar \tautau}

\newcommand{\pnfiggeneral}[5]{
\begin{figure}[#1]
{\centering \resizebox*{!}{#2}%
{#3} \par}
 
\caption{\label{#4}
#5
}
\end{figure}
}

\newcommand{\pnfig}[5]{
\pnfiggeneral{#1}{#2}{\includegraphics{#3}}{#4}{#5}
}

\def\Pythia{\textsc{Pythia}}

\def\Simdet{\textsc{Simdet}}

\def\Hdecay{\textsc{Hdecay}}

\def\ZBHT{\textsc{Zvtop-B-Hadron-Tagger}}

\begin{document}

\title{{\small{2005 International Linear Collider Workshop - Stanford,
U.S.A.}}\\ 
\vspace{12pt}
Extended analysis of the MSSM Higgs boson production at the Photon Collider} 

%

\author{P.\ Nie\.zurawski, A.\ F.\ \.Zarnecki}
\affiliation{Institute of Experimental Physics, Warsaw University, ul. Ho\.za 69, 00-681 Warsaw, Poland}
\author{M.\ Krawczyk}
\affiliation{Institute of Theoretical Physics, Warsaw University, ul. Ho\.za 69, 00-681 Warsaw, Poland}

\begin{abstract}

New analysis of the heavy, neutral MSSM Higgs bosons $H$ and $A$ production
at the Photon Collider is presented for \MAOeq 200, 250, 300 and 350 GeV
in the parameter range corresponding to the so called "LHC wedge" and beyond. 
The expected precision of the cross 
section measurement  for the process $\gagaAHbb$
and the "discovery reach" of the Photon Collider 
are compared for different MSSM scenarios.
The analysis takes into account all relevant theoretical and 
experimental issues which could affect the measurement.
For MSSM Higgs bosons $\AO$ and $\HO$, for \MAOeq 200--350~GeV and $\tbseven$, 
the statistical precision of the cross-section determination is 
estimated  to be 8--34\%, after one year of Photon Collider running,
for four considered MSSM parameters sets.
As heavy neutral Higgs bosons in this scenario may not be discovered at LHC 
or at the first stage of the $\epem$ collider, 
an opportunity of being a discovery machine is also studied for the Photon Collider.

\end{abstract}

\maketitle

\thispagestyle{fancy}

\section{INTRODUCTION} 

A photon-collider option of the future $\epem$ linear collider 
offers unique possibility to produce neutral Higgs bosons 
as $s$-channel resonances.
In case of the Minimal Supersymmetric extension of the Standard Model (MSSM) 
production of three neutral Higgs bosons, $h$, $\AO$ and $\HO$, can be considered.
For the light Higgs boson $h$ the statistical precision of the cross section measurement of about 2\%
is expected, similar to the Standard Model case \cite{SM_LCWS05_SLAC}.
In this contribution we estimate the precision of the corresponding measurement 
in case of the heavy MSSM Higgs bosons.
If  MSSM parameter values are in the so-called ``LHC wedge'', 
\ie region of intermediate values of $ \tan \beta$, $ \tan \beta \approx$ 4--10,
and masses $\MAOHO$ above 200~GeV,
the heavy pseudoscalar and scalar bosons, $\AO$ and $\HO$, may not be discovered 
at the LHC \cite{ATLAS_CMS_TDR,SearchATLAS,CMSDiscovery} 
and at the first stage of the $\epem$ linear collider \cite{TESLATDR_part3}.
Therefore we also study the measurement of the cross section $\gagaAHbb$ 
at the Photon Collider to evaluate the discovery potential of this experiment.
Possibility of discovering heavy MSSM Higgs bosons at the Photon Collider
has already been studied in the past (see \eg \cite{Gunion}). 
However, many relevant experimental aspects of the measurement are taken into account
in this study for the first time.
Parameter range considered in this analysis corresponds to 
a SM-like scenario 
where the lightest MSSM Higgs boson $h$ has properties similar to the SM Higgs boson, 
while heavy neutral Higgs bosons are nearly degenerated in mass 
and have negligible couplings to the gauge bosons $W/Z$. 
%


\section{MSSM SCENARIOS}

We consider MSSM scenarios described by parameter sets  similar to those used in \cite{MMuhlleitner},
\ie $ \tbseven , \, \mu = \pm 200 $~GeV  and $M_{2} = $ 200~GeV; 
these two parameter sets 
we denote as $I$ and $III$, see Tab.\ \ref{tab:MSSMparsets}. 
As compared to \cite{MMuhlleitner}, the values 
of trilinear couplings are changed (from  $A_{\widetilde{f}} = $ 0 to $A_{\widetilde{f}} = $ 1500~GeV),
so that the mass of the lightest Higgs boson, 
instead of being around 105~GeV (for \tanbeq 4 and \MAOeq 300~GeV)
is above the current lower limit for the SM Higgs boson mass, $\Mh > $ 114.4~GeV.
The intermediate scenario $II$ with $\mu = -150 $~GeV was also proposed.
For comparison with predictions presented by LHC experiments, the scenario $IV$
 used in \cite{CMSDiscovery} is also included.
The common sfermion mass equal to 1 TeV was assumed in all scenarios.
We have checked that all parameter sets imply masses of 
neutralinos, charginos, sleptons and squarks
higher than current experimantal limits.

\begin{table}[t]
\label{tab:MSSMparsets}
\bc
\begin{tabular}{|c|c|c|c|c|}
\hline
Symbol & $\mu$ [GeV] & $M_2$ [GeV] & $A_{\widetilde{f}}$ [GeV]  & $M_{\widetilde{f}}$ [GeV] \\
\hline
 $I$     &  200        &  200        &  1500                   &  1000           \\
 $II$    & -150        &  200        &  1500                   &  1000           \\
 $III$   & -200        &  200        &  1500                   &  1000           \\
 $IV$    &  300        &  200        &  2450                   &  1000           \\
\hline
\end{tabular}
\ec
\caption{MSSM parameter sets used in the described analysis.} 
\end{table}
%

%
Total widths and branching ratios of the Higgs bosons and the $\HO$ mass 
were calculated with the program \Hdecay{} \cite{HDECAY},
taking into account decays to and loops of supersymmetric particles.
These parameters were used during  generation of events 
with the \Pythia{} program \cite{PYTHIA}.
For the studied range of parameter values the heavy neutral Higgs bosons, $\AO$ and $\HO$,
are nearly mass degenerate.
The mass difference $\MHO - \MAO$  decreases with increasing $\tanb$ and $\MAO$
and is similar for all considered parameter sets.  
For  \MAOeq 200~GeV the mass difference decreases from
$\MHO - \MAO \approx 12 $~GeV for \tanbeq 3 to 0.7~GeV for \tanbeq 15,
whereas for \MAOeq 350~GeV the corresponding values are 6~GeV and 0.3~GeV, respectively.
The mass difference is larger or comparable 
to the total widths of  $\AO$ and $\HO$ which vary between 50~MeV and 4~GeV.
However, in most cases it is smaller than the invariant mass resolution, which is of
the order of 10~GeV.
Therefore, it is only possible to measure the total cross section for  $\AO$ and $\HO$ 
production.
In the considered parameter range
the branching ratios relevant for this study change between 3\% and 90\% for $\BrAOHObb$,
and from $2\cdot 10^{-7} $  to $ 9\cdot 10^{-5}$ for $\BrAOHOgaga$.
As processes $\gagaHbb$ and $\gagaAbb$ do not interfere,
the total $\gagaAHbb$ production rate is equal to the sum of both contributions.
Described Monte Carlo simulation of the heavy MSSM Higgs boson production at the Photon Collider 
was performed for scenario $I$ and parameter value  $ \tbseven $, and the obtained results 
were rescaled to other scenarios and the parameter range $\tanb = 3 - 20$.


\section{ANALYSIS}

 This analysis is based on the realistic simulations of the $\gaga$-luminosity spectra 
for the Photon Collider at TESLA \cite{V.TelnovSpectra,CompAZ}. 
It is assumed that the centre-of-mass energy of colliding electron beams, 
$\sqrt{s_{ee}}$, is optimized for the production of a Higgs bosons with a given mass.
Presented results  are obtained for a total integrated luminosity 
expected after one year of the Photon Collider running.
The distribution of the primary vertex 
and the beams crossing angle are taken into account.
As the main background to the Higgs-bosons production 
the heavy-quark pair production was considered;
the event samples were  generated using 
the program by G.~Jikia \cite{JikiaAndSoldner}
based on the NLO QCD results. 
Other background processes, which were neglected in the earlier analyses, 
were also studied:
$\gagaWW$, $\gagatautau$, and light-quark pair production $\gagaqq$.
Due to the large cross section and huge $\gaga$-luminosity at low $\Wgaga$,
about  two  \gagahad{} events (so-called overlaying events)
are expected   per bunch crossing.
To evaluate their  impact on the reconstruction of other events
produced in the same bunch crossing,
we generated \gagahad{} events with \Pythia{},
and overlaid them on signal and background events
according to the Poisson distribution.  
The detector performance was simulated
by the program   \Simdet{}  \cite{SIMDET401}.
Jets were reconstructed using the Durham algorithm.
The low-angle tracks and clusters were not taken into account 
to minimize the influence of \gagahad{} overlaying events.
Two or three jet events were accepted.
To reduce heavy-quark production background 
the lower cut on the polar angle for each jet 
and the upper cut on the total longitudinal momentum of the event were imposed.
Additional cuts to suppress $\gagaWW$ background were also applied.
For realistic \btagging{} the
\ZBHT{}  package was used \cite{HawkingBT,XellaBT,Btagging}.
The  criteria of event selection were optimized separately  for each considered Higgs-boson mass.
More detailed description of event generation, simulation and selection cuts
can be found in \cite{SM_LCWS05_SLAC} and \cite{PNThesis}.


\section{RESULTS}

The result of the analysis for \MAOeq 300~GeV is shown in Fig.\ \ref{fig:plot_var34_m300_modmssm_oe1_costhtc0.85}.
Distribution of the corrected invariant mass, $W_{corr} \equiv \sqrt{W_{rec}^{2}+2P_{T}(E+P_{T})}$ (see \cite{NZKhbbm120appb}), 
expected after one year of Photon Collider running, after imposing
all selection cuts is presented  for signal and all background contributions.
From the number of signal and background events in 
the optimized $W_{corr}$-window the expected statistical precision
of the cross-section measurement  is 11\%.

\pnfig{p}{\figheight}{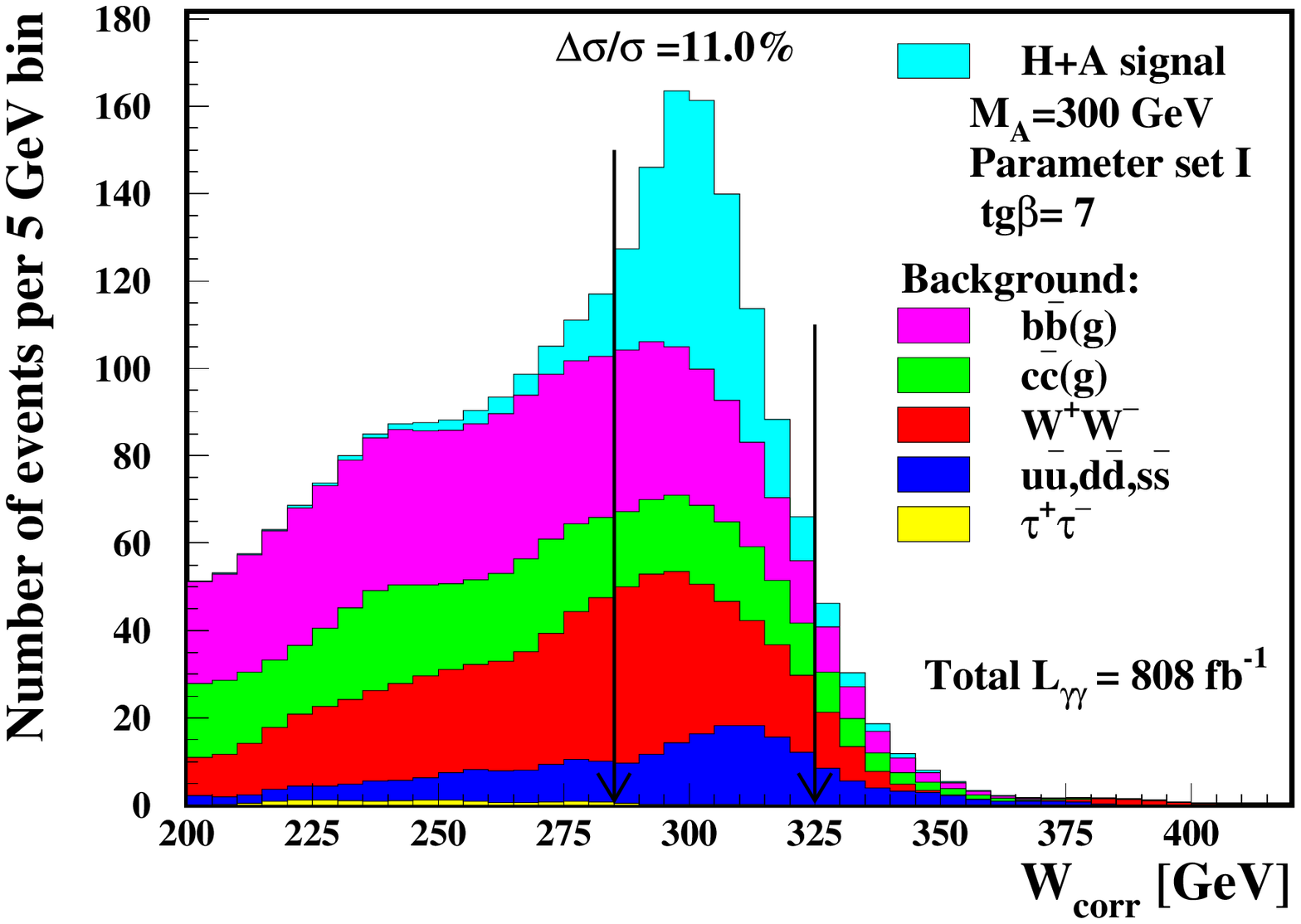}{fig:plot_var34_m300_modmssm_oe1_costhtc0.85}{
Distributions of the corrected invariant mass, $W_{corr}$, for signal and all considered background contributions,
with  overlaying events included.
The best precision of 11\% for $\gagaAHbb$ cross section measurement 
is achieved in the $W_{corr}$ window between   285 and  325~GeV.
}
\pnfig{p}{\figheight}{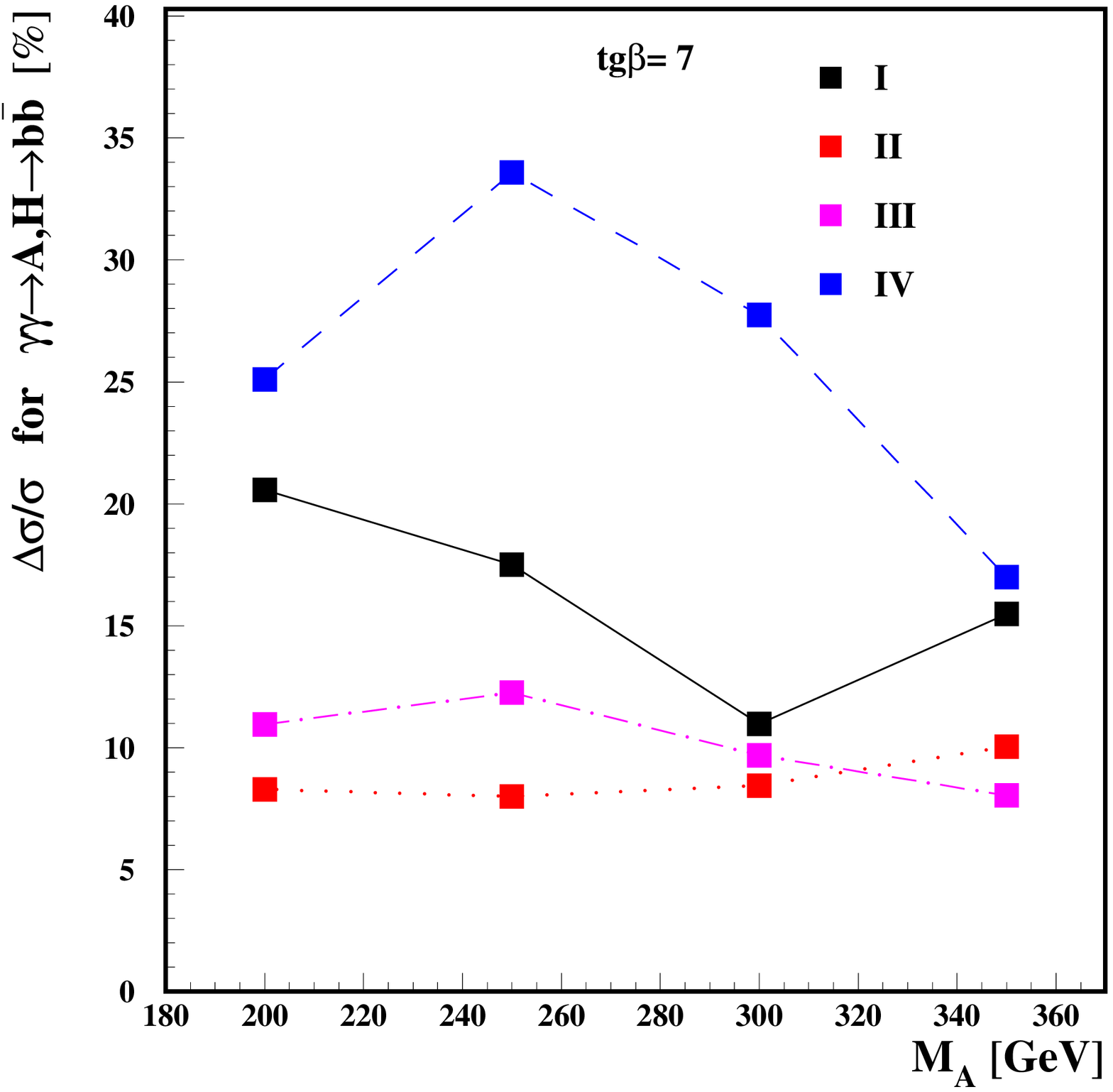}{fig:plot_mssm_precision_vs_ma_tgb7}{
Precisions of $\sgagaAHbb$ measurement are shown  for MSSM parameter sets $I$-$IV$, for \MAOeq 200--350~GeV and   \tanbeq 7.
The lines are drawn to guide the eye. 
}

We estimated statistical precision  for $\gagaAHbb$ cross section measurement
for all considered parameter sets.
Results obtained for   \tanbeq 7 are compared
in  Fig.\ \ref{fig:plot_mssm_precision_vs_ma_tgb7}.
The most precise measurement is expected  
for parameter sets $II$ and $III$  --- precision 
is about 10\% and
hardly depends on $\MAO$.
The worst measurement is expected for scenario $IV$, \ie the one considered by 
the CMS collaboration~\cite{CMSDiscovery}.
For all considered values of $\MAO$ the dependence of the measurement precision 
on $\tanb$ was studied;
the results for \MAOeq 200 and 350~GeV are  shown 
in Fig.\ \ref{fig:mssm_precision_vs_tgb_ma200_350}.
The precision  weakly  depends on  $\tanb$ 
if parameter sets $II$ or $III$ are  considered.
We also observed that for greater $\MAO$ values 
better precision of the cross section determination can be achieved.
In case of parameter sets  $I$ or $IV$ the precise measurement will not be possible
for low  $\tanb$ values,   $\tanb \lesssim 5 $.

\pnfiggeneral{t}{\twofigheight}{\includegraphics{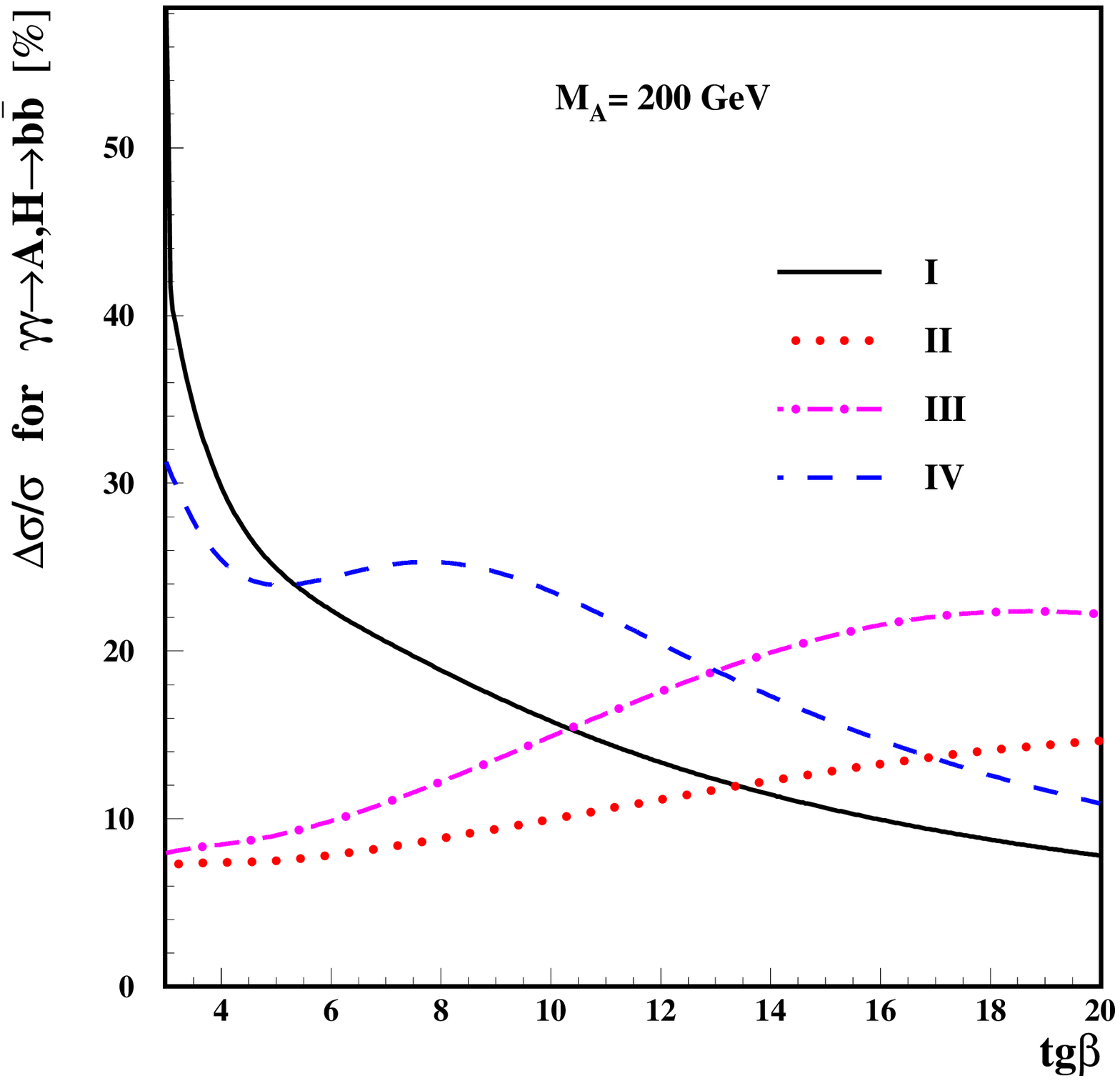}
\includegraphics{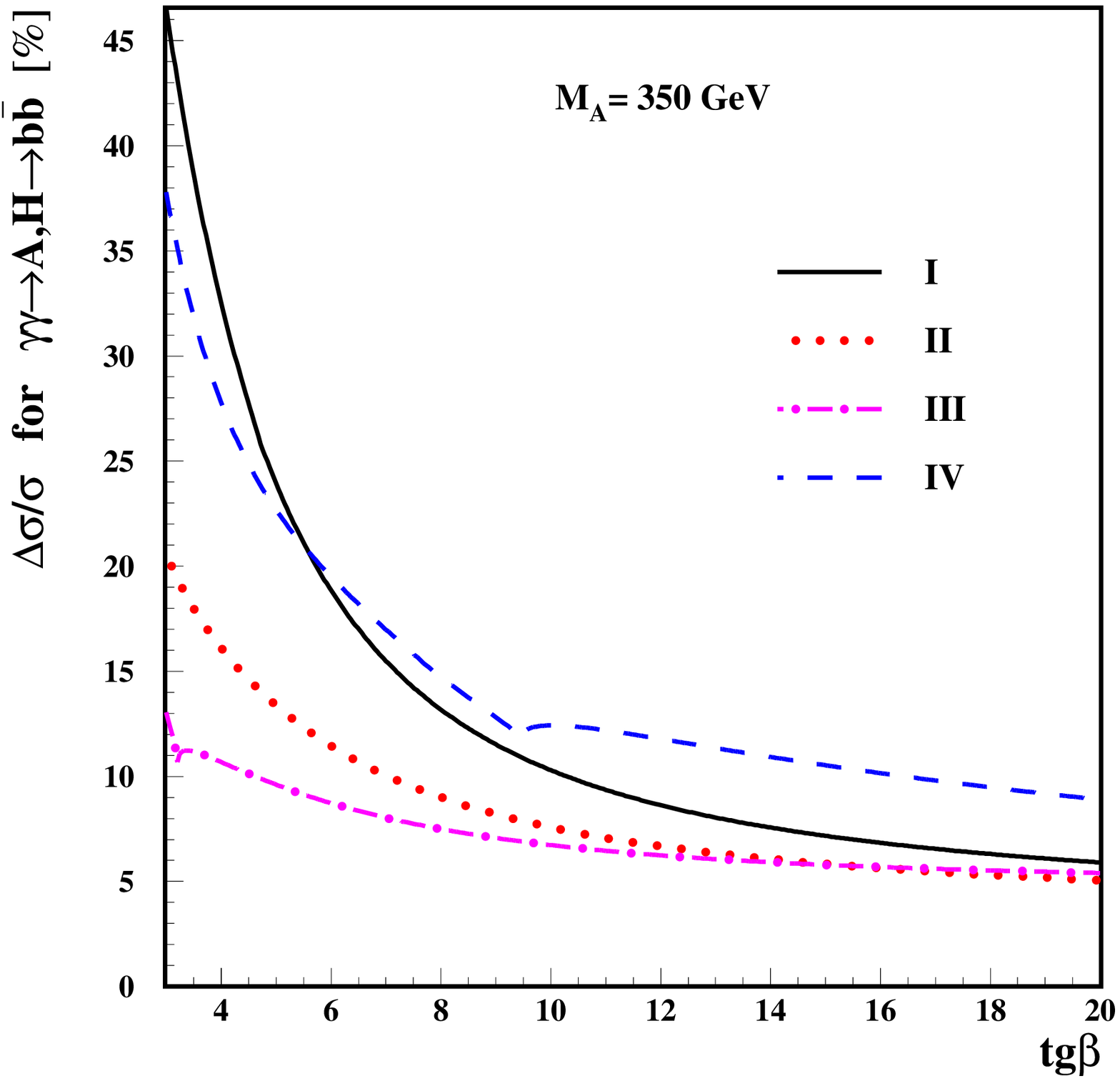} }{fig:mssm_precision_vs_tgb_ma200_350}{
Precisions of $\sgagaAHbb$ measurement are shown  for \MAOeq 200, 350~GeV, for MSSM parameter sets $I$-$IV$ with \tanbeq 3--20.
}

After discovery or a 'hint' of the resonant-like excess of events 
at LHC or ILC
the Photon Collider can be used
to confirm the observation and to measure the cross section
for production of the new state.
Thus, for all considered values of $\MAO$ we  studied  the significance
of  signal measurement 
as a function of  $\tanb$, for \tanbeq 3--20.
Results obtained for different parameter sets for \MAOeq 200, 350~GeV are compared 
in Fig.\ \ref{fig:mssm_significance_vs_tgb_ma200_350}.
The estimated lower limit of the discovery region of LHC experiments
(as presented by CMS collaboration \cite{CMSDiscovery}) is indicated by arrows.
For all parameter sets the expected statistics of signal events 
for \MAOeq 200--350~GeV will be sufficient 
to cover most of  the considered MSSM parameters space.
We can conclude that for $\MAO \gtrsim 300 $~GeV the Photon Collider 
should be able to discover Higgs bosons
for much lower values of $\tanb$ than experiments at the LHC.

\pnfiggeneral{tb}{\twofigheight}{\includegraphics{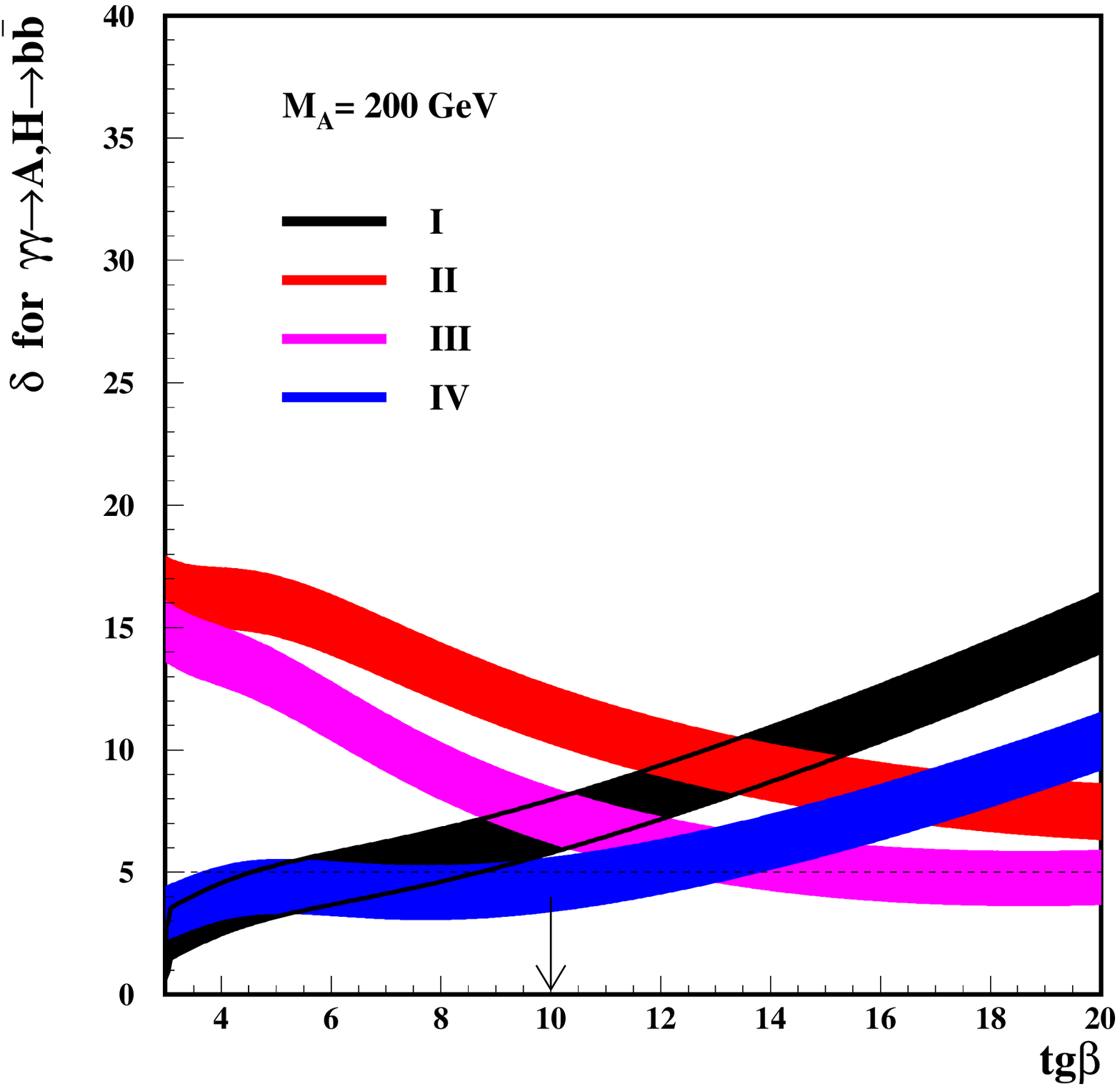}
\includegraphics{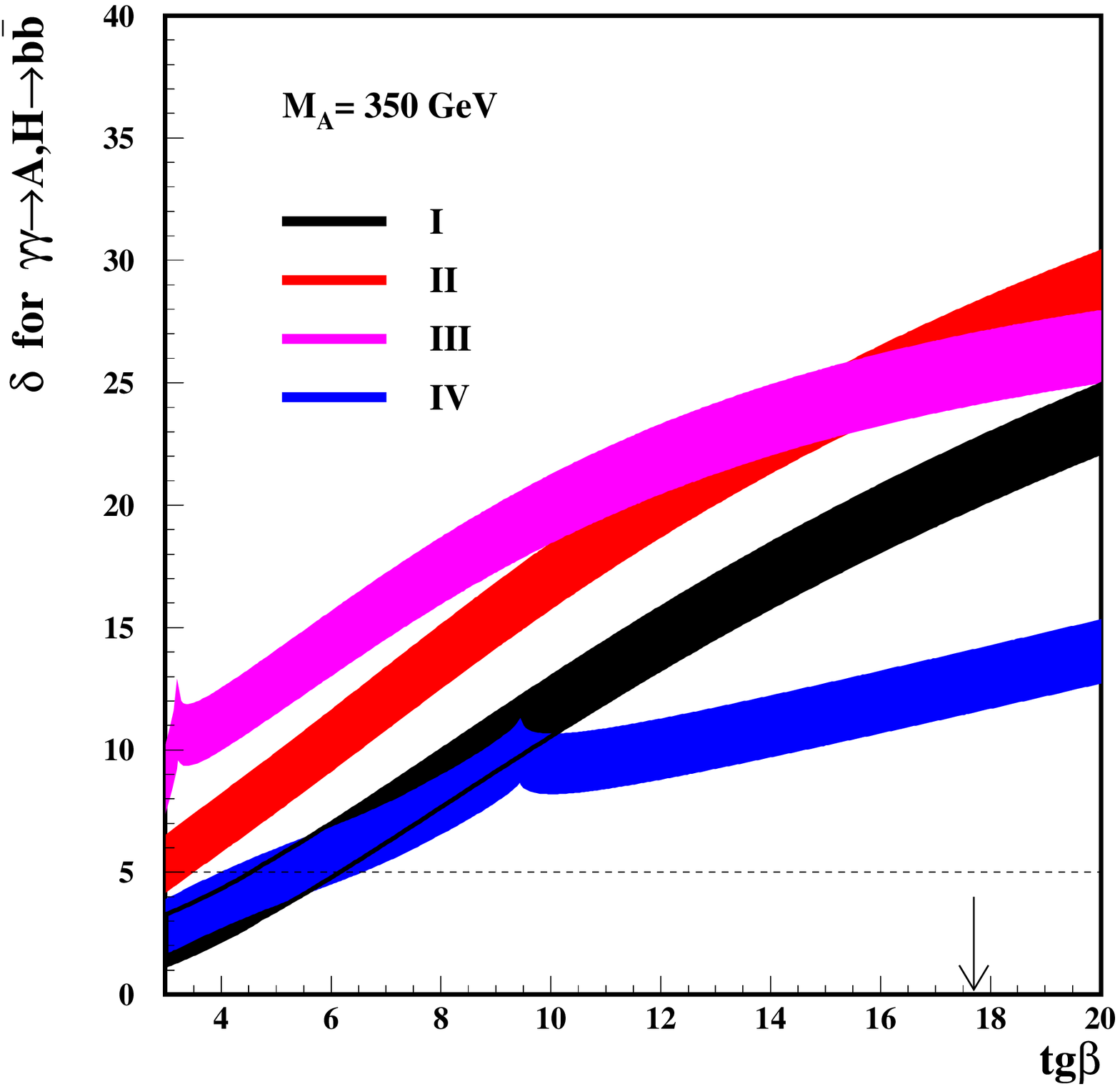} }{fig:mssm_significance_vs_tgb_ma200_350}{
Statistical significances of $\sgagaAHbb$ measurement are shown  for \MAOeq 200, 350~GeV, for MSSM parameter sets $I$-$IV$ with \tanbeq 3--20.
The band widths indicate the level of possible  statistical fluctuations of the actual measurement.
}

\section{SUMMARY}
In the presented analysis the production of heavy MSSM Higgs bosons
$\AO$ and $\HO$ at the Photon Collider, in the process $\gagaAHbb$,
was studied 
taking into account all relevant experimental and theoretical aspects.
The considered MSSM parameter range  corresponds to the so-called  ``LHC wedge'',
where identification of $\AO$ and $\HO$ may not be possible at LHC and $\epem$ collider.
Our analysis shows that, for  $\MAO \sim$ 300~GeV,
the cross section for the MSSM Higgs-bosons production $\sgagaAHbb$ can be measured
with a statistical precision of about 11\% already after one year of Photon Collider
running at nominal luminosity.
For other considered values of $\MAO$   it turns out to be  lower -- from 16\% to 21\%.
Although this result is less optimistic than the earlier estimate \cite{MMuhlleitner},
  still the photon--photon collider gives opportunity
of a precision measurement of $\sgagaAHbb$, 
assuming that we know the mass of the Higgs boson(s). 
A discovery of  MSSM Higgs-bosons requires energy scanning or a run with 
a broad luminosity spectrum, perhaps  followed by the run with a peaked one. 
We estimate the  significance  expected for the Higgs production measurement 
for four different  parameter sets and for \tanbeq 3--20.
The discovery reach of the Photon Collider is compared with the estimated reach of the CMS experiment.
 For the optimum energy and polarizations choice the photon--photon collisions allow for the discovery of the Higgs bosons 
even for $\tanb$ values  lower than the expected reach of the LHC experiments.
Thus, at least partially, the Photon Collider will cover the so-called ``LHC wedge''.
For low $\tanb$ values the measurement 
could probably profit from use of additional channels: 
decays of $\AO$ and $\HO$ to charginos and neutralinos,
and in case of $\HO$ also decays to $hh$.

\clearpage

\begin{acknowledgments}
We would like to thank M.~M.~M\"{u}hlleitner for valuable discussions.
We also acknowledge useful comments and suggestions 
of other colleagues from the ECFA/DESY working groups.
This work was partially supported 
by the Polish Committee for Scientific Research, 
grant  no.~1~P03B~040~26
and
project no.~115/E-343/SPB/DESY/P-03/DWM517/2003-2005.
P.N.~acknowledges partial
support by Polish Committee for Scientific Research, grant 
no.~2~P03B~128~25.
\end{acknowledgments}



\begin{thebibliography}{9}   

\bibitem{SM_LCWS05_SLAC}
P. \Niezurawski, Contribution No.~0503 in these Proceedings.

\bibitem{ATLAS_CMS_TDR}
ATLAS Coll., Technical Design Report, CERN-LHCC 99-14 (1999). \\
CMS Coll., Technical Proposal, CERN-LHCC 94-38 (1994).

\bibitem{SearchATLAS} S.\ Gentile, ATL-PHYS-2004-009.

\bibitem{CMSDiscovery} S.\ Abdullin \etal, CMS NOTE 2003/033.

\bibitem{TESLATDR_part3}
  J.~A.~Aguilar-Saavedra  \etal, %
  hep-ph/0106315.


\bibitem{Gunion}
D.M.~Asner, J.B.~Gronberg, J.F.~Gunion,
Phys.\ Rev.\ D~67 (2003) 035009, hep-ph/0110320.


\bibitem{MMuhlleitner} 
M.~M.~M\"uhlleitner, M.~Kr\"amer, M.~Spira, P.~M.~Zerwas, 
Phys. Lett. B 508 (2001) 311, hep-ph/0101083. 


\bibitem{HDECAY}A.~Djouadi, J.~Kalinowski, M.~Spira, Comput.~Phys.~Commun.~108 (1998) 56, \mbox{hep-ph/9704448}.

\bibitem{PYTHIA}T.~Sj\"ostrand \etal, 
Comput.~Phys.~Commun.~135 (2001) 238, hep-ph/0108264.

\bibitem{V.TelnovSpectra}
\mbox{V.\ I.~Telnov}, 
http://www.desy.de/\textasciitilde{}telnov/ggtesla/spectra/.

\bibitem{CompAZ}
A.\ F.\ \Zarnecki, 
Acta Phys.~Polon.~B 34 (2003) 2741, hep-ex/0207021. 

\bibitem{JikiaAndSoldner} G.~Jikia, S.~S\"oldner-Rembold, 
Nucl.~Instrum.~Meth.~A 472 (2001) 133, hep-ex/0101056. 


\bibitem{SIMDET401}M.~Pohl, H.~J.~Schreiber,
DESY-02-061, hep-ex/0206009.

 
\bibitem{HawkingBT}
R.\ Hawkings,
LC-PHSM-2000-021-TESLA.

\bibitem{XellaBT}
S.~M.~Xella Hansen, D.~J.~Jackson, R.\ Hawkings, C.~Damerell,
LC-PHSM-2001-024.

\bibitem{Btagging}T.~Kuhl, K.~Harder, 
talk presented at the II Workshop of ECFA-DESY Study,
Saint~Malo, April 2002. 
%


\bibitem{PNThesis}
P. \Niezurawski, hep-ph/0503295.

\bibitem{NZKhbbm120appb}
P. \Niezurawski, A.F. \Zarnecki, M. Krawczyk,  
Acta Phys.\ Polon.\ B 34 (2003) 177,
\mbox{hep-ph/0208234}. 



\end{thebibliography}
\end{document}